\documentclass[journal]{IEEEtran}
\usepackage{amsmath,amsfonts,bbm}
\usepackage{algorithm}
\usepackage{algorithmic}
\usepackage{array}
\usepackage[caption=false,font=normalsize,labelfont=sf,textfont=sf]{subfig}
\usepackage{cite}
\usepackage{textcomp}
\usepackage{stfloats}
\usepackage{url}
\usepackage{verbatim}
\usepackage{graphicx}
\usepackage{stfloats}
\usepackage{color}
\usepackage{enumitem}
\def\BibTeX{{\rm B\kern-.05em{\sc i\kern-.025em b}\kern-.08em
    T\kern-.1667em\lower.7ex\hbox{E}\kern-.125emX}}
\usepackage{balance}

\usepackage[normalem]{ulem}

\begin{document}
\title{Integrated Sensing, Communication and Control enabled Agile UAV Swarm}
\author{Zhiqing Wei,~\IEEEmembership{Member,~IEEE},
Yucong Du,~\IEEEmembership{Student Member,~IEEE}, 
Zhiyong Feng,~\IEEEmembership{Senior Member,~IEEE},\\
Haotian Liu,~\IEEEmembership{Student Member,~IEEE}, 
Yanpeng Cui,~\IEEEmembership{Student Member,~IEEE},
Tao Zhang,\\
Ying Zhou,~\IEEEmembership{Student Member,~IEEE},
Huici Wu,~\IEEEmembership{Member,~IEEE}
\thanks{This work was funded
in part by Fundamental and Interdisciplinary Disciplines Breakthrough Plan of the Ministry of Education of China under Grant JYB2025XDXM119,
in part by Beijing Natural Science Foundation under Grant Z220004,
in part by BUPT-China Unicom Joint Innovation Center,
in part by the National Natural Science Foundation of China (NSFC) under Grant 62271081,
in part by the Fundamental Research Funds for the Central Universities under Grant 2024ZCJH01,
and in part by the National Key R\&D Program of China under Grant 2023YFA1009600.

Z. Wei, Y. Du, Z. Feng, H. Liu, and Y. Zhou are with 
the Key Laboratory of Universal Wireless Communications, Ministry of Education, 
Beijing University of Posts and Telecommunications, Beijing 100876, China 
(email: \{weizhiqing, duyc, fengzy, haotian\_liu, zhouying9705, dailywu\}@bupt.edu.cn).

Y. Cui is with School of Information Engineering, Henan Institute of Science and Technology, 
Xinxiang 453003, Henan, China (email: cuiyp@hist.edu.cn).

T. Zhang is with China Unicom Research Institute, Beijing 100176‌, China (email: zhangt91@chinaunicom.cn).

H. Wu is with the National Engineering Research Center of Mobile Network Technologies, 
Beijing University of Posts and Telecommunications, Beijing 100876, China, 
and also with Pengcheng Laboratory, Shenzhen 518066, China (e-mail: dailywu@bupt.edu.cn).}}
\markboth{}
{}
\maketitle

\begin{abstract}
Uncrewed aerial vehicle (UAV) swarms are pivotal in the applications 
such as disaster relief,
aerial base station (BS) and logistics transportation.
These scenarios require the capabilities in accurate sensing,
efficient communication and flexible control
for real-time and reliable task execution.
However, sensing, communication and control are studied independently in traditional research,
which limits the overall performance of UAV swarms.
To overcome this disadvantage, we propose a deeply coupled scheme of
integrated sensing, communication and control (ISCC) for UAV swarms,
which is a systemic paradigm that transcends traditional
isolated designs of sensing, communication and control by
establishing a tightly-coupled closed-loop through
the co-optimization of sensing, communication and control.
In this article, we firstly analyze the requirements of
scenarios and key performance metrics.
Subsequently, the enabling technologies are proposed,
including communication-and-control-enhanced sensing,
sensing-and-control-enhanced communication,
and sensing-and-communication-enhanced control.
Simulation results validate the performance of the proposed ISCC framework,
demonstrating its application potential in the future.
\end{abstract}


\section{Introduction}
Uncrewed aerial vehicle (UAV) swarms have played an indispensable role in the scenarios
such as disaster relief, aerial base station (BS) and logistics transportation.
In the disaster relief scenario,
UAV swarms can be quickly deployed to perform collaborative tasks including remote sensing of affected areas,
locating trapped individuals and monitoring environmental threats,
while sharing sensing information relying on efficient communication systems.
In the aerial BS scenario,
UAVs detect user distribution and channel characteristics,
fuse environmental information, and optimize the positions and resources of UAVs
to support large-scale user access.
In the logistics transportation scenario,
UAVs leverage accurate sensing and stable communication to understand complex environment
and receive scheduling commands,
enabling efficient collision avoidance and swarm control.
In summary, UAV swarms need to have the capabilities in accurate sensing,
efficient communication and flexible control in complex dynamic environment
to ensure the safe, real-time and reliable task execution.

However, in traditional UAV swarms,
the separated design of sensing, communication and control functions results in 
substantial protocol interaction delay and heavy processing overhead between these functions.
Consequently, the sensing, communication and control systems of UAV swarms
cannot efficiently exchange collaborative information,
leading to the inability to rapidly adapt to the complex environment.
These challenges are promising to be solved by
the seamless integration of sensing, communication and control,
which is promising to construct agile UAV swarms,
enabling the widespread application of UAV swarms in complex environment
under limited payload and resource constraints.

In recent years, the rise of the low-altitude economy (LAE)
explores the application of UAV swarms in complex environment,
where their sensing, communication
and control capabilities are essential for efficient task execution.
Current research in academia and industry mainly focuses on
integrated sensing and communication (ISAC)-enabled UAV swarms
and trajectory control for UAV swarms.
\begin{itemize}
    \def\labelenumi{\arabic{enumi})}
    \item \textbf{ISAC-enabled UAV swarms:}
        The ISAC-enabled UAV swarms are currently addressed in terms of architecture,
        resource allocation and networking.
        These studies are all conducted under the constraints of the size, weight and power (SWaP) 
        limitations inherent to UAV platforms,
        with the ultimate goal of enabling intelligent and collaborative task execution.
        \textit{For UAV network architecture},
        Wu et al. \cite{1} systematically analyzed the
        key challenges and future opportunities for ISAC-enabled UAV networks.
        Meanwhile, Hazarika et al. \cite{4} proposed an ISAC network framework and protocol,
        which significantly enhances system robustness 
        in dynamic environments by incorporating spatiotemporal cluster analysis and the Fréchet distance algorithm.
        \textit{For resource allocation},
        Wu et al. \cite{1} demonstrated that ISAC functions can substantially
        improve both data sharing efficiency and environmental sensing.
        Meng et al. \cite{3} developed a joint trajectory and resource allocation scheme for UAV swarms,
        aiming to joint optimize communication and sensing performance metrics.
        Zhang et al. \cite{5} investigated the application of ISAC within UAV swarms,
        focusing on enhancing real-time situational awareness and
        adapting communications to meet dynamic task requirements.
        \textit{For UAV networking},
        Mu et al. \cite{2} conducted a forward-looking review,
        identifying critical challenges and future research directions for ISAC-empowered UAVs, 
        with a specific focus on the application of ISAC between UAV and BS,
        as well as UAV-assisted sensing for ground targets.
        Zhang et al. \cite{5} introduced an adaptive topology reconstruction strategy,
        which effectively improves both sensing efficiency and 
        operational endurance of UAVs in dynamic environment.

    \item \textbf{UAV trajectory control:}
        High-accuracy trajectory control is essential 
        for enhancing the task execution efficiency and reliability of UAV swarms.
        Current researches include but not limited to
        path planning and communication performance optimization.
        \textit{For path planning},
        Wu et al. \cite{6} addressed the challenge of rapid UAV deployment in dynamic environments
        by developing an optimization framework that applies
        real-time sensing to dynamically adjust flight trajectories.
        Li et al. \cite{7} presented a UAV data distribution system with expected data requirements.
        This system aims for energy efficiency by employing strategies
        such as power allocation during hovering and curvature-adaptive trajectory adjustment.
        \textit{For communication optimization},
        Wu et al. \cite{6} focused on trajectory design to deal with the throughput-delay-energy trade-off,
        reducing energy consumption while maintaining communication performance.
        Li et al. \cite{7} investigated energy-efficient operations by 
        restructuring the balance between task duration and propulsion energy 
        via joint power allocation and adaptive path planning.
        Hu et al. \cite{8} leveraged reinforcement learning to
        manage dynamic topology and channel uncertainty in cellular-connected UAVs,
        proposing a protocol that improves the efficiency of sensing information collection.
\end{itemize}

Overall, current studies on ISAC-enhanced UAV swarms and 
control of UAV swarms are independent.
The disconnect between sensing, communication and control
ultimately limits the potential application of UAV swarms in complex environment.
Hence,
we propose a framework of integrated sensing, communication and control (ISCC) 
to construct agile UAV swarms.
ISCC integrates sensing, communication and control functions,
forming a multifunctional wireless network infrastructure.
Empowered by multimodal sensors and ISAC technologies,
the functions of sensing, communication and control in ISCC share hardware and resources,
realizing complementary enhancement among sensing, communication and control.
ISCC optimizes the performance of sensing, communication and control under limited resources,
improving resource utilization and overall efficiency of UAV swarms.
ISCC offers a systematic solution for agile UAV swarms in complex and dynamic environment.

The remainder of this article is organized as follows.
The scenarios, requirements and performance metrics are described in Section II.
We introduce the application of ISCC in the scenarios of disaster relief, 
aerial BS and logistics transportation in the subsequent three sections.
Then, the performance of the proposed enabling technologies is evaluated.
Finally, this article is summarized.

\section{Scenarios, requirements and performance metrics}

\begin{figure*}
	\centering
	\includegraphics[width=1\linewidth]{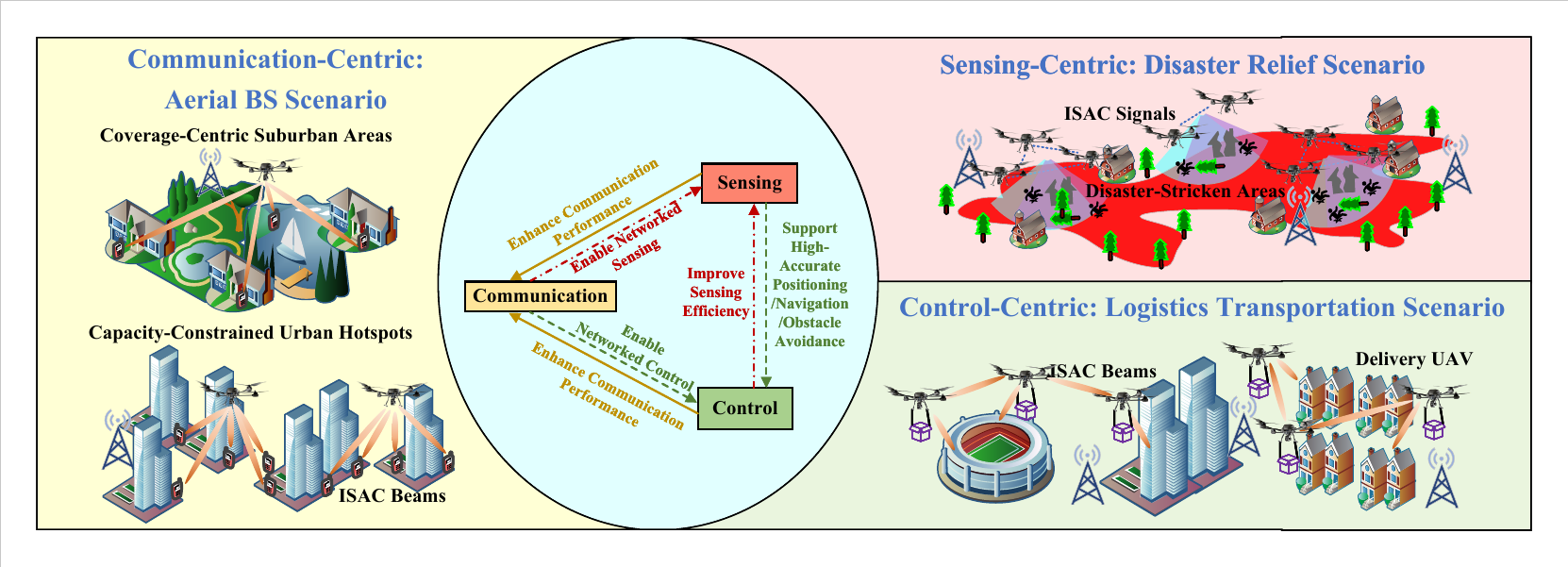}
	\caption{Scenarios of ISCC UAV swarms.}
	\label{fig1}
\end{figure*}

In this section, the scenarios and requirements for disaster relief,
aerial BS and logistics transportation are firstly introduced.
Then, the performance metrics of sensing, communication and control are introduced.
As shown in Fig. \ref{fig1},
through the complementary enhancement of sensing, communication and control,
the performance of UAV swarms is enhanced.

\subsection{Scenarios and Requirements}

\subsubsection{Disaster Relief}

This scenario focuses on the requirement of
rapid environmental sensing,
enabling rapid exploration, mapping and survivor detection,
where communication and control are leveraged to assist the core sensing function.
With efficient communication enabling sensing information sharing,
UAVs are able to extend the sensing range and improve sensing accuracy.
With location and trajectory control of UAVs, 
the performance of sensing can be optimized on demand.

\begin{enumerate}[label=\alph*)]
\item \textbf{Scenario Description and Requirements:}

    \quad \textbf{Rapid Situation Assessment:}
    UAVs are quickly deployed in disaster-stricken areas to perform disaster assessment and survivor searching.

    \quad \textbf{Accurate Localization of Survivors:}
    Utilizing ISAC payloads and multi-modal sensors is crucial for 
    accurate localization of survivors,
    while also facilitating the identification of obstacles and environmental threats.

\item \textbf{Specific Advantages of ISCC UAV Swarms:}

    \quad In the disaster relief scenario,
    the ISCC framework endows UAV swarms with flexible sensing capabilities.
    Specifically, communication allows UAVs to share sensing information in real time, 
    thereby enhancing sensing accuracy and expanding coverage range. 
    Meanwhile, control enables UAVs to achieve on-demand sensing coverage of disaster-stricken areas.
\end{enumerate}

\subsubsection{Aerial BS}

This scenario focuses on coverage-centric suburban areas and capacity-constrained urban hotspots, 
where sensing and control are leveraged to optimize the core communication performance.

\begin{enumerate}[label=\alph*)]
\item \textbf{Scenario Description and Requirements:}

    \quad \textbf{Dynamic Coverage Adaptability Requirement:}
    The aerial BS should be capable of autonomously and continuously 
    adjusting its coverage to match the dynamic distribution of ground users, 
    a capability that is essential for maximizing service availability.

    \quad \textbf{Fast Networking Requirement:}
    Fast inter-UAV and UAV-to-ground networking schemes are essential 
    to adapt to the dynamic, uncharted and complex environment.

\item \textbf{Specific Advantages of ISCC UAV Swarms:}
    \quad In the aerial BS scenario,
    the ISCC framework empowers UAVs to provide flexible wireless coverage 
    by using real-time sensing for proactive channel estimation and beamforming toward user-dense areas,
    while adaptively controlling UAV positions to sustain stable communication links.
\end{enumerate}

\subsubsection{Logistics Transportation}

This scenario focuses on achieving reliable autonomous logistics delivery
in complex airspace via robust navigation and trajectory control strategies,
where ISCC framework is leveraged to ensure the safe, efficient and collision-free control of UAVs.

\begin{enumerate}[label=\alph*)]
\item \textbf{Scenario Description and Requirements:}

    \quad \textbf{Emergency-Aware Autonomous UAV Control:}
    The UAVs need to be capable of autonomously sensing and
    reacting to unexpected obstacles or dynamic threats
    such as birds and unauthorized UAVs in real time.
    This necessitates an immediate and stable control response to collision avoidance,
    guaranteeing flight safety.

    \quad \textbf{Resilient Path Planning:}
    The UAVs need to be endowed with resilient navigation and path planning capabilities to operate in complex environments, 
    including urban canyons and building interiors.

\item \textbf{Specific Advantages of ISCC UAV Swarms:}
    \quad In the logistics transportation scenario, 
    the ISCC framework empowers UAVs to realize adaptive autonomous flight control.
    Specifically, this framework provides real-time environment sensing,
    which in turn enables precise path planning and 
    facilitates rapid collision avoidance in response to unforeseen obstacles.
\end{enumerate}

In the following subsections,
we introduce the key performance metrics of sensing, 
communication and control.

\subsection{Performance Metrics}

\subsubsection{Communication Performance Metrics}
Communication performance metrics such as channel capacity, neighbor discovery accuracy and routing table update time are introduced in this subsection. 

\begin{enumerate}[label=\alph*)]
\item \textbf{Channel Capacity:}
Channel capacity represents the maximum data rate that can be reliably sustained over the communication link. 
The value of channel capacity is determined by the bandwidth and the 
signal-to-interference-plus-noise ratio (SINR).

\item \textbf{Neighbor Discovery Accuracy and Routing Table Update Time:}
Neighbor discovery accuracy is defined as the consistency between the neighbor set identified by a node 
and the set of nodes actually reachable within its communication range. 
Routing table update time refers to the time interval from the occurrence of 
a network topology change to the completion of routing information correction by all relevant nodes.
\end{enumerate}

\subsubsection{Sensing Performance Metrics}

Sensing performance metrics are mainly categorized into detection and estimation performance metrics.
Detection performance is typically measured using the probability of detection and the probability of false alarm,
while estimation performance is commonly evaluated via 
the minimum mean squared error (MMSE) and the Cramér-Rao lower bound (CRLB).

\begin{enumerate}[label=\alph*)]
\item \textbf{Probability of Detection:}
The probability of detection is defined as the likelihood that a system correctly 
confirms the presence of a target when the target actually exists.

\item \textbf{Probability of False Alarm:}
The probability of false alarm is defined as the likelihood that a system erroneously
declares the presence of a target when no target actually exists.

\item \textbf{Minimum Mean Squared Error:}
The MMSE is a performance metric for evaluating the accuracy of parameter estimation, 
which quantifies the average squared difference between the real value of a parameter and its estimated value.
Besides, MMSE metric extends to the average root mean square error (ARMSE) metric for multi-target sensing,
which is expressed as the average of the roots of the mean square errors (MSEs) corresponding to multiple targets.

\item \textbf{Cramér-Rao Lower Bound:}
For an unbiased parameter estimator,
the variance of its estimation must be greater than or equal to the value of the CRLB. 
As a fundamental performance benchmark for evaluating estimation algorithms, 
the CRLB is derived from the inverse of the Fisher information matrix (FIM).
\end{enumerate}

\subsubsection{Control Performance Metrics}

This article focus on the performance of path planning and collision avoidance, 
including path planning time, energy consumption 
and collision probability.

\begin{enumerate}[label=\alph*)]
\item \textbf{Path Planning Time:}
Path planning time is defined as the total duration required for UAVs to generate feasible 
and optimal flight paths from their current locations to the destinations.

\item \textbf{Energy Consumption:}
Energy consumption is defined as the total energy consumed by UAVs 
throughout their entire flight paths \cite{9}. 

\item \textbf{Collision Probability:}
Collision probability is defined as the probability that a UAV collides 
with obstacles during flight,
which is associated with the distance between the UAV and the obstacles 
as well as the UAV's sensing accuracy \cite{10}. 
\end{enumerate}

The following section introduces three enabling technologies for UAV swarms, 
namely communication-and-control-enhanced sensing, 
sensing-and-control-enhanced communication, 
and sensing-and-communication-enhanced control.

\section{Communication-and-Control-Enhanced Sensing for UAV Swarms} \label{sensing}
\begin{figure}[t]
	\centering
 	\subfloat[]{
		\includegraphics[width=0.5\linewidth]{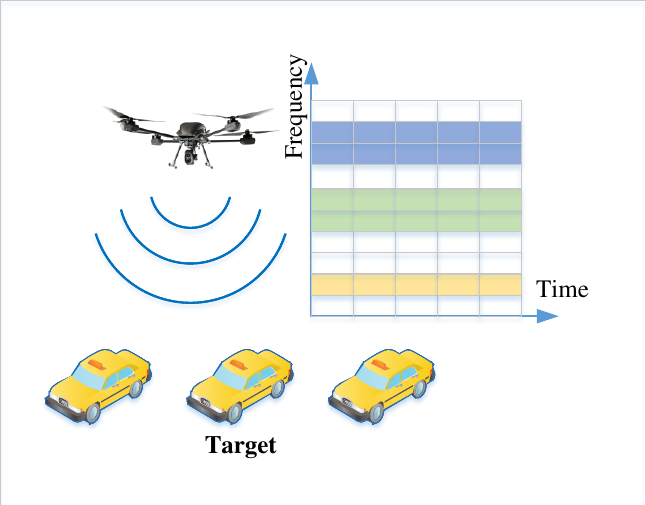} 
		\label{f8a} 
	}
    \centering
 	\subfloat[]{
		\includegraphics[width=0.5\linewidth]{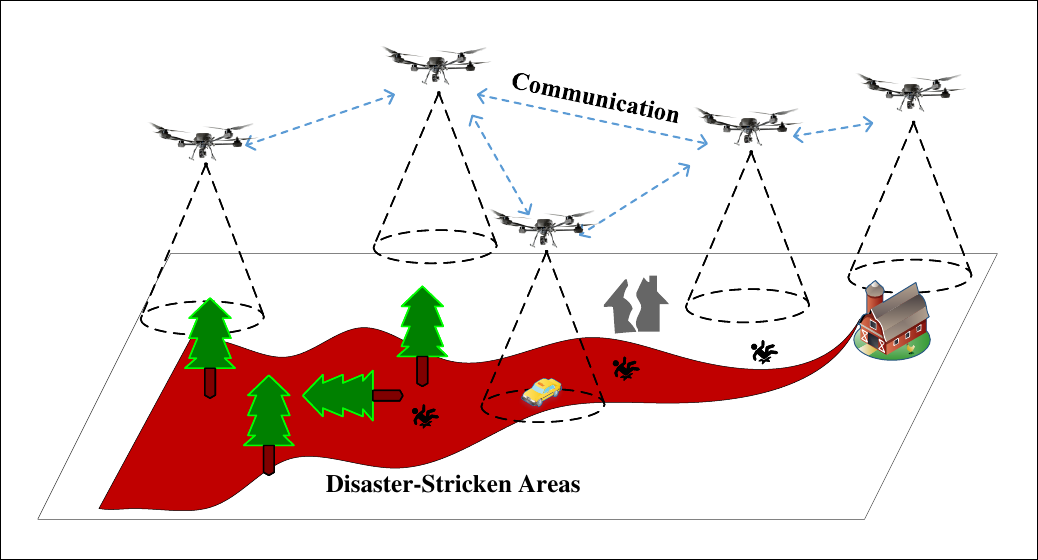} 
		\label{f8b} 
	}
     \quad
	\subfloat[]{
		\includegraphics[width=0.7\linewidth]{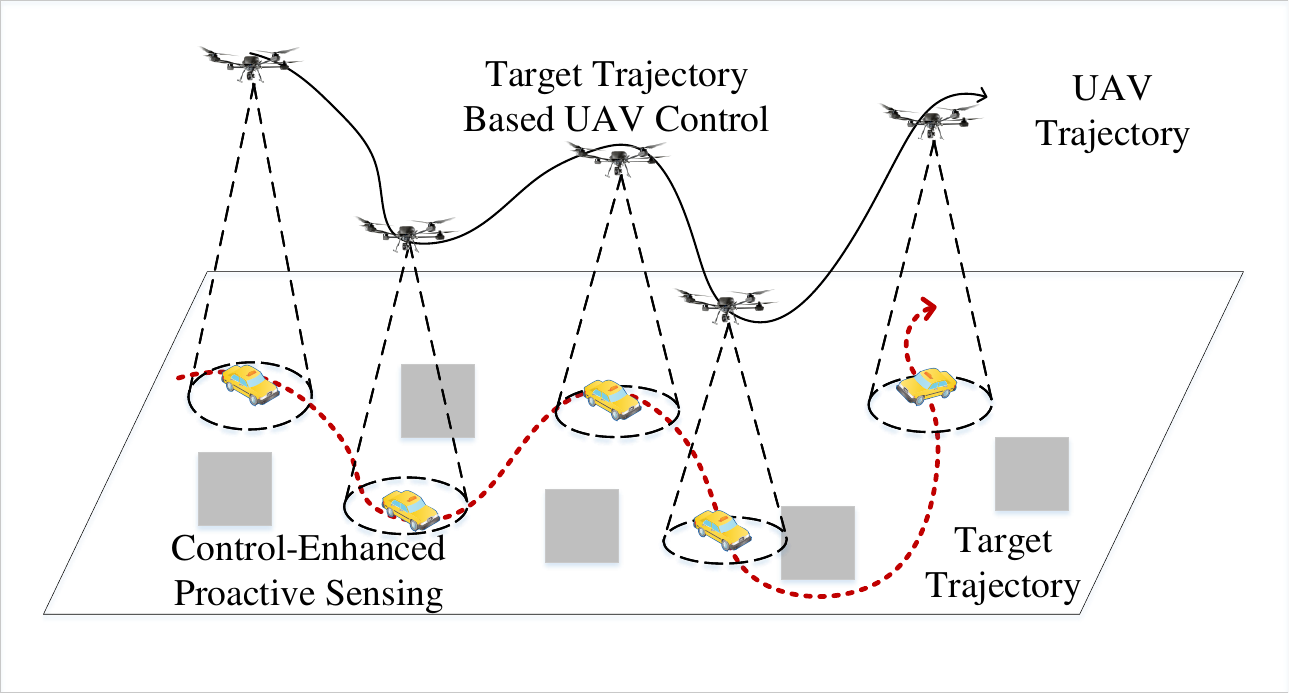}
		\label{f8c} 
	}
	\caption{Communication-and-control-enhanced sensing for disaster relief scenario:
                    (a) the communication-enhanced sensing signal design;
                    (b) the communication-enhanced sensing signal processing;
                    (c) the control-enhanced sensing for UAV swarms.}
	\label{fig8}
\end{figure}

In this section, the enabling technologies for enhancing sensing performance are proposed, 
including communication-enhanced sensing signal design, 
communication-enhanced sensing signal processing,
and control-enhanced sensing for UAV swarms,
as shown in Fig. \ref{fig8}.

\subsection{Communication-Enhanced Sensing Signal Design}

The bandwidth available to UAV swarms is becoming increasingly limited and fragmented.
Therefore, it is difficult for the sensing functions to obtain a continuous and usable bandwidth,
thereby constraining further performance improvements.
A feasible solution, building upon mature communication architectures, 
is to first detect idle bands through spectrum sensing and 
then apply carrier aggregation to generate dynamic multi-band sensing signals.
However, the core challenge lies in how to effectively
leverage these non-contiguous spectrum resources to achieve high-accuracy sensing.

To address this challenge, a blank-band recovery method based on the all-pole model can be applied~\cite{allpole}.
Specifically, the sensing echo signal derived from the orthogonal frequency division multiplexing (OFDM)-based 
ISAC waveform can be formulated as an all-pole model,
which in turn enables the representation of the echo signal via the poles and amplitudes in this model.
On this basis, we leverage observation data that includes blank data to estimate the model's poles and amplitudes,
and subsequently utilize these estimated parameters to reconstruct the blank data.
Finally, by iteratively repeating the aforementioned process,
the blank frequency bands can be recovered and denoised,
thereby maximizing the utilization of fragmented spectrum resources.
This approach ensures robust sensing performance even in spectrum-congested disaster relief scenarios.
However, the efficacy of this method hinges on breakthroughs in key enabling technologies, 
such as high-precision rapid spectrum sensing,
efficient dynamic resource allocation,
and low-complexity model recovery algorithms.

\subsection{Communication-Enhanced Sensing Signal Processing}
For effective survivor localization in disaster-stricken areas,
high-accuracy sensing of multiple ground targets is crucial. 
However, complex factors commonly present in disaster relief scenarios,
such as widespread non-line-of-sight (NLoS) propagation,
the sensitive target radar cross-section (RCS)
and dense target distributions,
posing significant challenges for sensing,
especially for eliminating sensing blind zones and resolving multiple targets.
To overcome the inherent limitations of single-viewpoint sensing,
we propose a multi-UAV cooperative sensing method
that utilizes inter-UAV communication links 
to facilitate multi-view information sharing and cooperative sensing.

The core of this method resides in the design of a multi-strategy fusion algorithm,
which enables the deep integration of multi-view information
and thus effectively mitigates sensing degradation caused by adverse viewing angles or target occlusions.
This capability is essential for high-accuracy sensing in complex disaster-stricken areas.
Although ISAC cooperative sensing can substantially improve multi-target sensing accuracy,
relying solely on ISAC-based sensing makes it challenging to achieve robust target feature recognition and 
high-accuracy sensing under severe weather conditions. 
A promising solution lies in multi-modal sensing that fuses ISAC sensing information with visual data. 
However, several open challenges remain,
including the spatiotemporal alignment and feature fusion of multi-modal heterogeneous data. 
Furthermore,
the adoption of advanced technologies,
such as quantum technology \cite{R2},
holds enormous potential for boosting sensing performance.

\subsection{Control-Enhanced Sensing for UAV Swarms}

Efficient and full-coverage sensing in disaster-stricken areas requires 
dynamic UAV control rather than static or pre-programmed flight paths. 
With the ISCC framework, 
control is strategically elevated from its conventional role of mere mobility management
to a proactive approach that directly enhances sensing performance.

Specifically, this control mechanism proactively guides
UAV swarms to explore unknown or priority areas,
thereby maximizing sensing efficiency while avoiding redundant coverage and ineffective sensing operations. 
It also precisely coordinates the relative positions and sensor orientations of UAVs,
so as to enhance the sensing accuracy for disaster-stricken areas.

\section{Sensing-and-Control-Enhanced Communication for UAV Swarms} \label{communication}
\begin{figure*}
	\centering
	\includegraphics[width=1\linewidth]{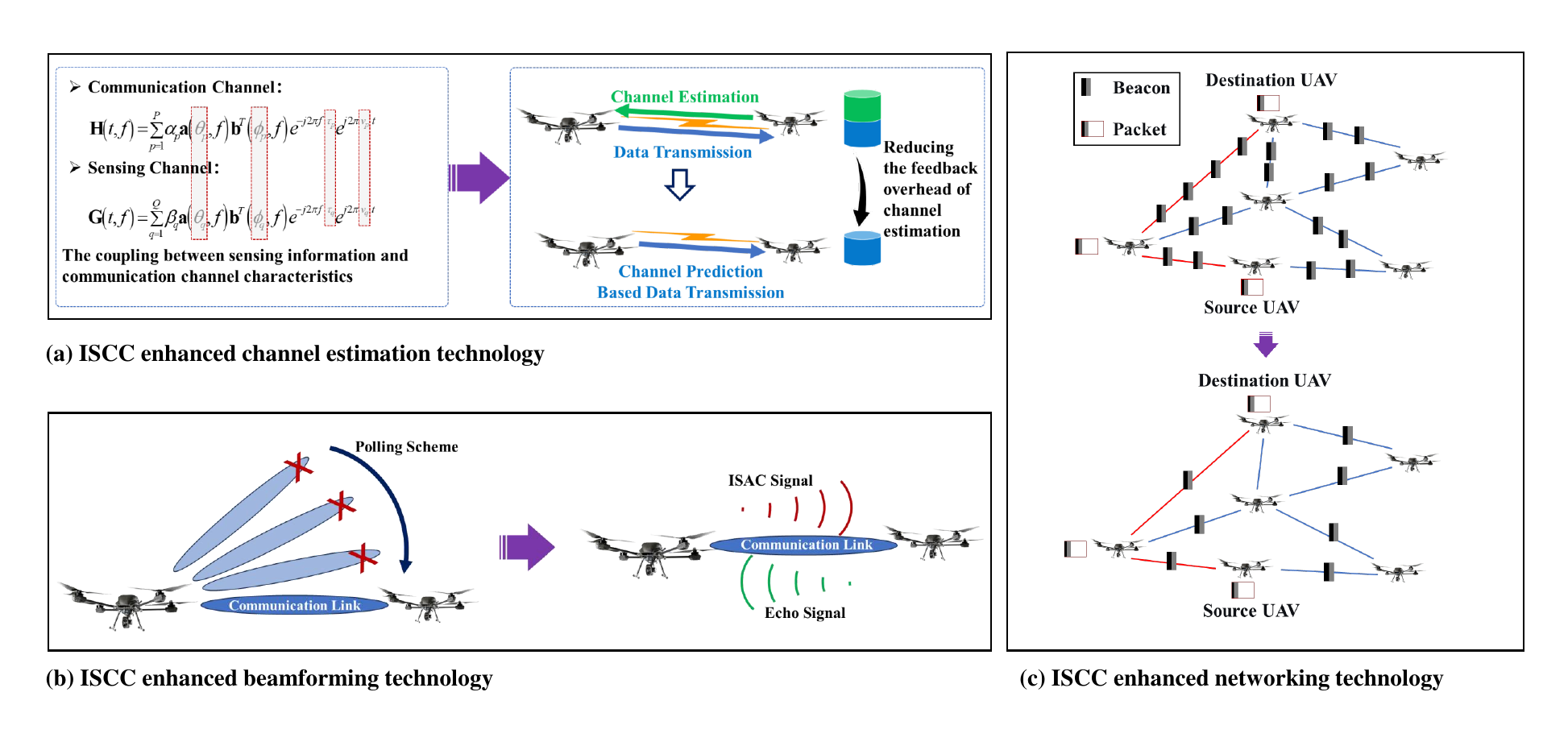}
	\caption{Sensing-and-control-enhanced communication for aerial BS scenario.}
	\label{fig2}
\end{figure*}

This section introduces three key enabling technologies supporting communication in dynamic environments,
namely ISCC-based channel estimation and beamforming technologies for
maintaining reliable links under time-varying channel conditions,
neighbor discovery and routing technologies for adapting to dynamic topology changes, 
and multiple access methods for efficiently allocating resources among UAVs,
as illustrated in Fig. \ref{fig2}.

\subsection{ISCC-Based Channel Estimation and Beamforming}
For aerial BS scenarios, traditional transmission techniques relying
solely on communication feedback incur substantial communication overhead.
This challenge is exacerbated in high-mobility urban environments,
where the positions of users are constantly changing.
To address this critical issue,
we propose ISCC-based channel estimation and beamforming schemes.

The ISCC-based channel estimation method improves upon the traditional feedback-driven channel estimation approach 
by harnessing sensing capabilities to obtain real-time angle-of-arrival (AoA) information of users as prior information.
As shown in Fig. \ref{fig2}a,
instead of relying on frequent pilot signals and excessive feedback overhead, 
this method introduces a proactive channel estimation framework, 
which is a critical design for maintaining link quality in rapidly time-varying wireless networks.
Furthermore, a model-driven deep learning scheme could be utilized, 
which refines channel gain estimation by integrating angle-based prior information with a residual network (ResNet). 
This approach not only effectively suppresses noise in received signals
but also achieves high-accuracy channel estimation with minimal training costs, 
thus ensuring the robust and efficient acquisition of channel state information (CSI).

ISCC-based predictive beamforming leverages accurate CSI and UAV echo signals to realize real-time beam alignment, 
thus obviating the need for frequent pilot signals and excessive feedback overhead.
As shown in Fig. \ref{fig2}b, 
by fusing sensing-derived physical identifiers with communication-based digital identifiers, 
this scheme enables accurate beam steering toward the users in dense multi-user scenarios.
Enhanced by an extended Kalman filter (EKF),
this approach maintains continuous and high-accuracy beam tracking even under rapidly time-varying channel environment.
This approach yields substantial gains in spectrum efficiency and transmission reliability 
while minimizing communication overhead, 
thereby fulfilling the core operational requirements of aerial BS in resource-constrained scenario.

\subsection{ISCC-Based Neighbor Discovery and Routing}
To sustain robust aerial connectivity,
rapid and continuous neighbor discovery is indispensable for highly dynamic UAV swarms leveraging ISAC signals.
By jointly processing sensing and communication data,
UAVs dynamically update the neighbor states (e.g., position and velocity), 
thereby ensuring adaptive and stable networking for UAV swarms.

This robust neighbor discovery capability directly facilitates efficient routing in UAV networks.
Traditional routing methods rely on the repeated exchange of network status information
among nodes to determine transmission paths,
which are inefficient and bring additional delay.
This inefficiency arises from the limitation that conventional networks only recognize the logical addresses of devices, 
rather than sensing their actual positions and motion states.
To address this limitation, the proposed sensing-assisted routing technology
fuses real-time data streams from wireless communication modules and onboard sensors, 
dynamically associating network identifiers with the physical features of UAVs 
(e.g., real-time location, velocity, color, appearance, etc.), 
as shown in Fig. \ref{fig2}c.
This technology enables UAVs to directly select optimal neighbor nodes,
thus substantially cutting down the communication overhead associated with network maintenance.

\subsection{ISCC-Based Multiple Access Methods}
The efficient networking of UAV swarms relies on advanced multiple access technology.
The performance of this technology is primarily characterized by key performance metrics 
including access delay, packet collision probability and network congestion probability,
which directly influence the quality of service (QoS)
in both coverage-centric suburban areas and capacity-constrained urban hotspots.
To sustain networking efficiency,
we propose a sensing-assisted coordinated resource allocation approach across the time, space and code domains. 
In this approach, real-time sensing capabilities are leveraged to
acquire the positions and velocities of neighboring nodes.
These data are then utilized to dynamically adapt beam directions, 
multi-domain resource allocation and access timings,
thus effectively mitigating packet collision risks.
This multi-domain resource coordination plays a pivotal role 
in maximizing spectrum efficiency of highly congested aerial networks.

\section{Sensing-and-Communication-Enhanced Control for UAV Swarms}\label{control}

\begin{figure*}
	\centering
	\includegraphics[width=0.7\linewidth]{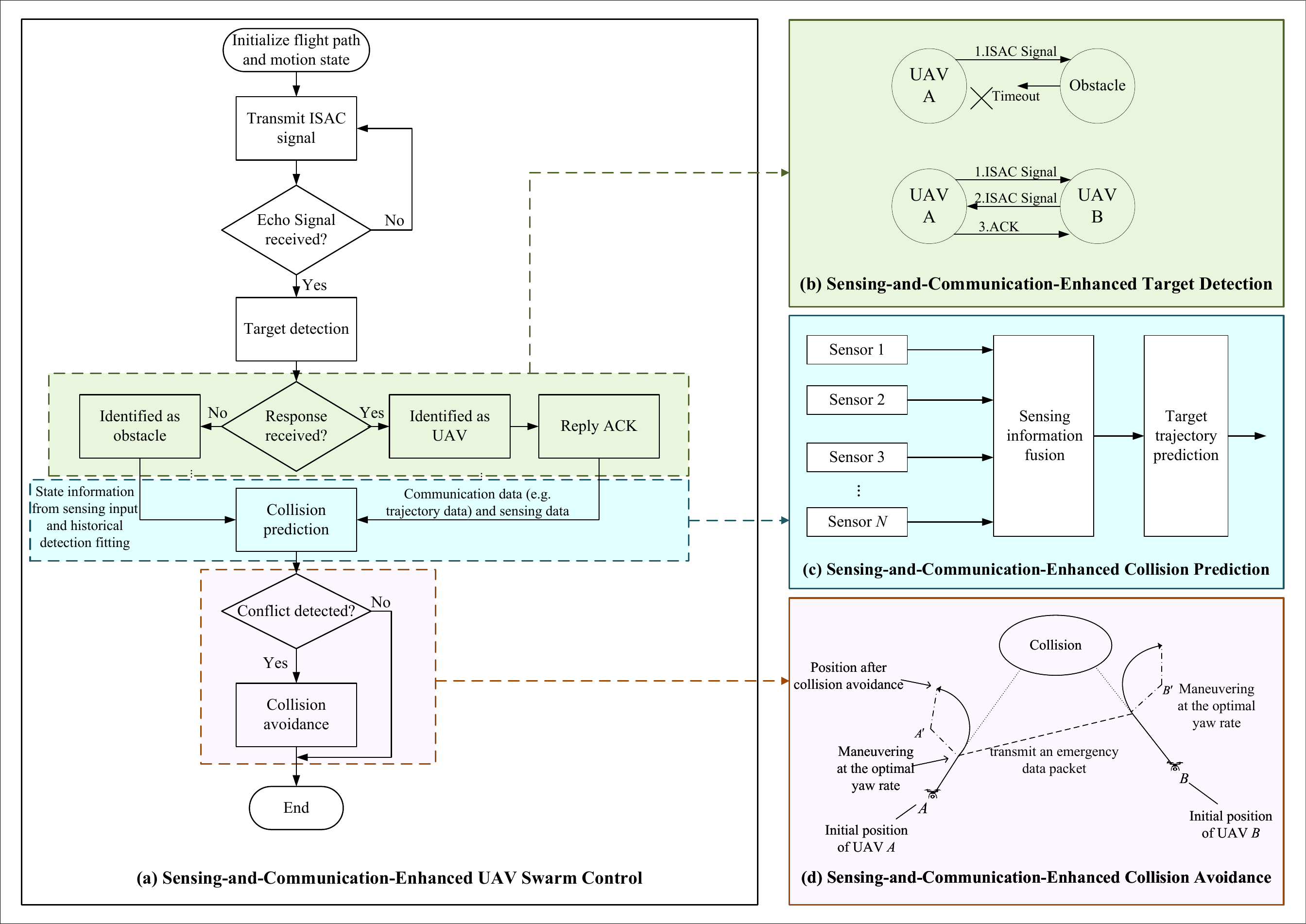}
	\caption{Sensing-and-communication-enhanced control for logistics transportation scenario.}
	\label{fig3}
\end{figure*}

This section presents the enabling technologies
tailored to the logistics transportation scenario,
where UAVs undertake point-to-point parcel delivery tasks 
that demand real-time collision avoidance capabilities.
The ISCC enhanced UAV swarm control is shown in Fig. \ref{fig3}a,
which includes three critical parts:
target detection,
collision prediction,
and collision avoidance.
All of them leverage ISCC framework to realize accurate, adaptive, 
and safe motion control of UAV swarms,
as shown in Fig. \ref{fig3}a.

\subsection{Sensing-and-Communication-Enhanced Target Detection} 
Target detection is the first step in UAV collision avoidance.
In logistics transportation scenarios,
the detection targets for UAVs include other UAVs and environmental obstacles.
ISAC is promising in realizing rapid target detection and recognition.
The ISAC-based UAV target detection process is shown in Fig. \ref{fig3}b. 
Specifically, UAV A first employs ISAC signals to perform a full scan of the environment.
If UAV A detects a target and does not receive any communication feedback from it, 
this target is categorized as an obstacle.
In this case, UAV A will only process the corresponding echo signal
to derive the target's location and velocity parameters.
Conversely, if the detected target sends a communication feedback to UAV A,
it is identified as another UAV (designated as UAV B).
In this scenario, UAV A will simultaneously capture both the echo signal and the communication feedback from UAV B. 
Hence, UAV A is able to realize target detection and recognition using ISAC signals.
On this basis, UAV A initiates collision risk prediction.
If the detected target is UAV B and no collision risk is identified, 
UAV A sends an acknowledgment (ACK) confirmation signal to UAV B.
Otherwise, UAV A sends an emergency data packet,
triggering both UAV A and UAV B 
to execute real-time collision avoidance maneuvers.

\subsection{Sensing-and-Communication-Enhanced Collision Prediction}
UAV initiates collision risk prediction to evaluate the urgency of obstacle avoidance.
It is noted that the communication range commonly exceeds the sensing range and 
the collision prediction method is based on the target's real-time location.
If the target falls within the sensing range,
the UAV leverages ISAC-based target detection and recognition.
On this basis, the UAV employs an EKF to fuse sensing information extracted from ISAC signals,
thereby enabling accurate target trajectory prediction.
If the target lies beyond the sensing range,
the UAV exchanges its trajectory data with another UAV via communication links and
predicts the collision risk based on this shared mutual trajectory data.
Furthermore, to achieve higher prediction accuracy,
the UAV swarm collaboratively fuses the shared sensing information 
prior to employing an EKF for collision prediction,
as illustrated in Fig. \ref{fig3}c.

\subsection{Sensing-and-Communication-Enhanced Collision Avoidance}
Upon completing target detection and collision prediction,
a UAV is required to formulate collision avoidance decisions immediately upon identifying a collision threat.
Given that variations in the shapes and sizes of UAVs have a direct impact on the collision model, 
a circumscribed sphere is adopted as a standardized collision model for both UAVs and obstacles.
The radius of this sphere is defined as the equivalent radius of the collision model,
which is a parameter that is dependent on the UAV's sensing accuracy.
Notably, a higher level of sensing accuracy corresponds to a smaller equivalent radius.
The minimum safe separation distance required for collision avoidance 
between a UAV and a target is determined by three key components,
namely the equivalent radii of the two collision models,
the response time of the UAV's braking system,
and the trajectory prediction error.
If the prediction result indicates an impending collision, i.e., 
the minimum safe separation distance cannot be ensured, 
the UAV will execute an emergency collision avoidance maneuver. 
In this case, if the object is an obstacle,
the UAV can evade it by maneuvering at the optimal yaw rate.
If the detected object is another UAV,
the UAV will transmit an emergency data packet that notifies the collision status and
the designated collision avoidance schemes. 
Subsequently, two UAVs shall initiate an emergency evasion maneuver at the optimal yaw rate, 
as illustrated in Fig. \ref{fig3}d.

\section{Performance Evaluation}

Using complementary enhancement of sensing, communication and control, 
this section presents three parts of performance evaluation,
including communication-and-control-enhanced sensing,
sensing-and-control-enhanced communication,
and communication-and-sensing-enhanced control.

\subsection{Communication-and-Control-Enhanced Sensing}

We use the range estimation performance of multiple targets to 
evaluate the blank frequency band recovery method based on 
the all-pole model in Section~\ref{sensing}.
The simulation parameters are configured as follows.
\begin{itemize}
    \item Carrier frequency: 24 GHz.
    \item Subcarrier spacing: 120 kHz.
    \item Total bandwidth: 61.44 MHz.
    \item Frequency gap: 30.72 MHz.
    \item Number of targets: 3 (ranges randomly distributed within the interval [200, 1000] m).
\end{itemize}
Two benchmark methods are considered:
an improved fast Fourier transform (FFT) method \cite{11} and 
the state-of-the-art (SOTA) compressed sensing method \cite{11}.
As shown in Fig. \ref{f5s1},
in low signal-to-noise ratio (SNR) regimes,
the proposed method incorporating truncated singular value decomposition exhibits noise susceptibility, 
resulting in the performance inferior to that of the benchmark methods.
As SNR increases, the proposed method demonstrates superior performance 
in frequency gap data recovery and higher accuracy in range estimation,
attributed to the prior physical constraints of its all-pole model.

\subsection{Sensing-and-Control-Enhanced Communication}

\begin{figure}[t]
	\centering
 	\subfloat[]{
		\includegraphics[width=0.7\linewidth]{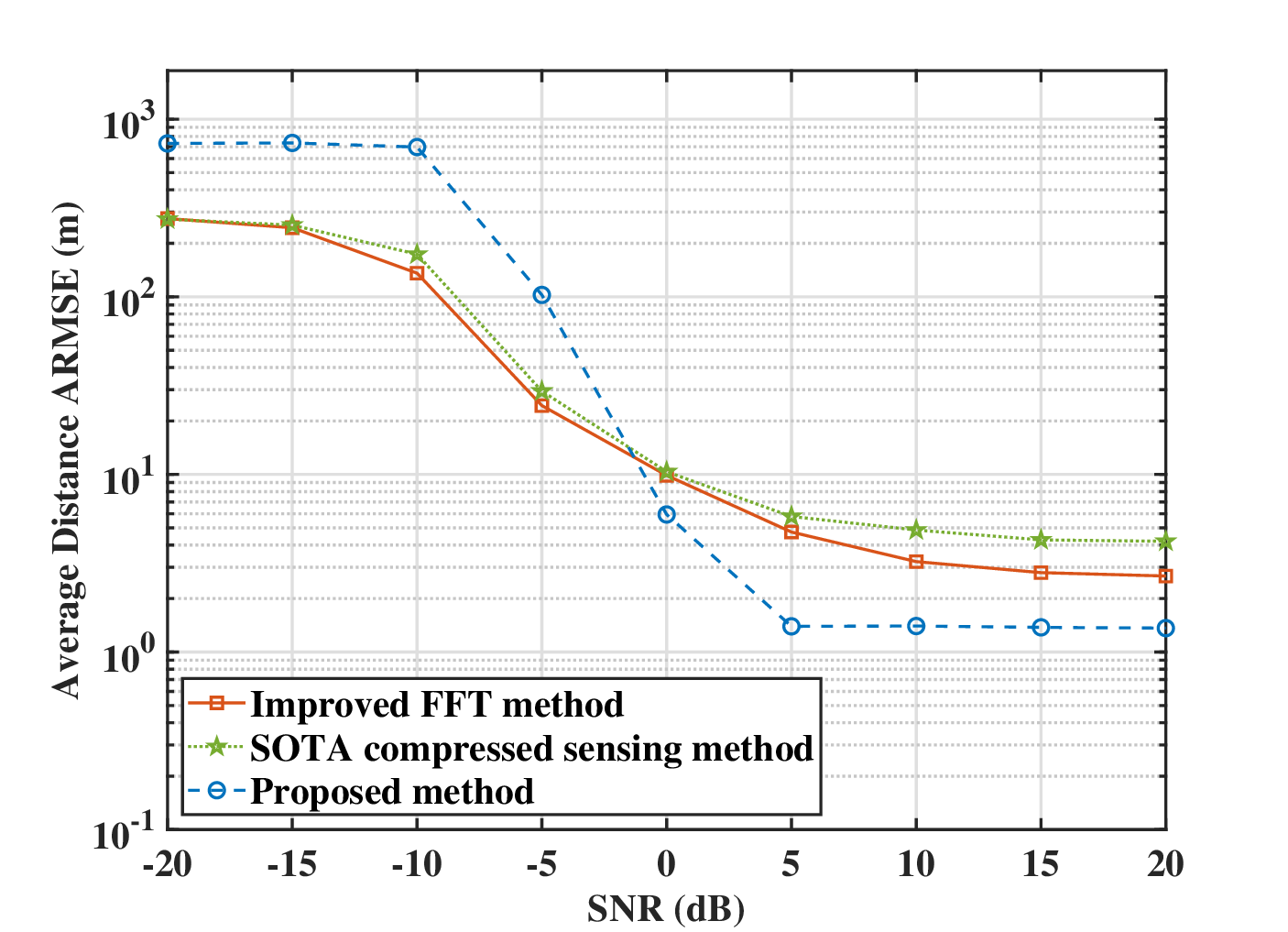} 
		\label{f5s1} 
	}
    \quad
    \centering
 	\subfloat[]{
		\includegraphics[width=0.5\linewidth]{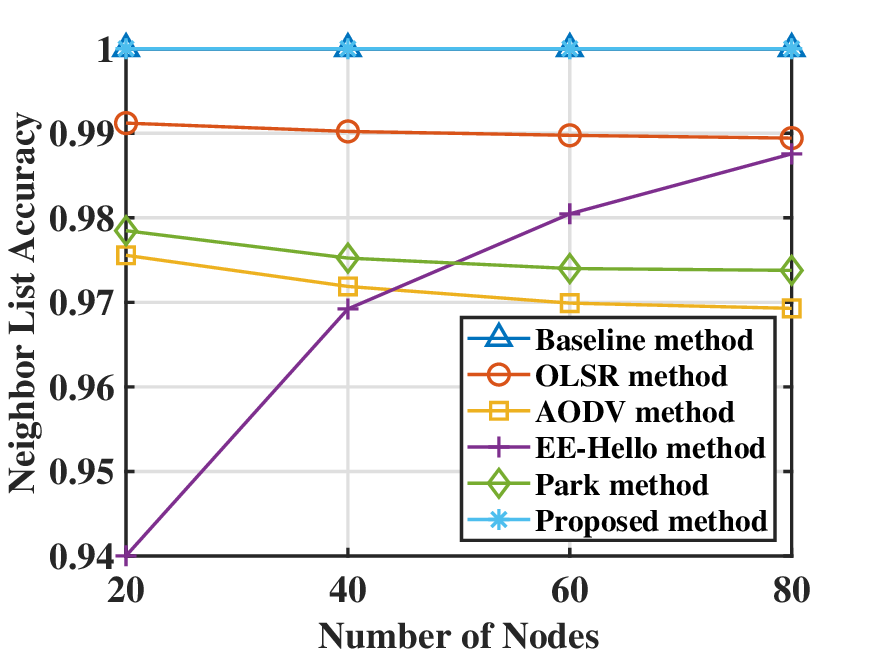} 
		\label{f5s2} 
	}
	\subfloat[]{
		\includegraphics[width=0.5\linewidth]{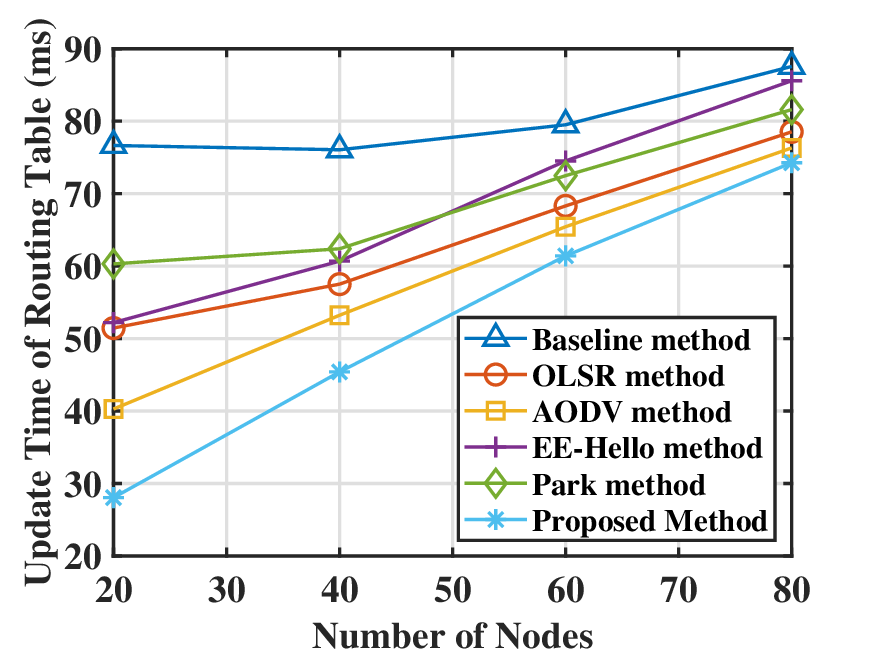}
		\label{f5s3} 
	}
	\caption{Performance evaluation of UAV Swarms:
                    (a) the ARMSE of communication-and-control-enhanced sensing for UAV Swarms;
                    (b) the accuracy of neighbor list in ISCC-enabled neighbor discovery;
                    (c) the update time of routing table in ISCC-enabled routing.}
	\label{fig5}
\end{figure}

We use the neighbor list accuracy and routing table update time 
to evaluate the networking performance based on the ISCC framework 
in Section~\ref{communication}.
The simulation parameters are configured as follows.
\begin{itemize}
    \item Communication range: 156 m.
    \item Transmit power: 1 W.
    \item SINR threshold: -14 dB.
    \item Velocity of UAV: 5-10 m/s.
    \item Number of UAVs: 20-80 (ranges randomly distributed within the three-dimensional (3D) interval [600, 600, 300] m).
\end{itemize}
Five benchmark methods are considered: 
0.25-second fixed beacon interval maintain method (baseline method), 
optimized link state routing (OLSR) method \cite{13}, 
ad hoc on-demand distance vector (AODV) method \cite{12}, 
energy efficient (EE)-Hello method \cite{14}, 
and Park method \cite{15}.

The evaluation of the neighbor discovery algorithm is shown
in Fig. \ref{f5s2}.
As the number of nodes increases,
the accuracy of the neighbor list of the EE-Hello \cite{14} method continues to improve,
while the OLSR method \cite{13}, 
the method proposed by Park et al. \cite{15}, 
and the AODV method \cite{12} experience a decline in neighbor discovery accuracy.
In contrast,
both the proposed method and the baseline method with
fixed 0.25-second beacon interval maintain
100\% neighbor accuracy
regardless of the increase in node numbers. 
The proposed method eliminates the need for frequent beacon transmissions. 
The beacon exchanges are triggered only when nodes detects that
they enter into or exit from the effective communication range by sensing. 
Consequently, it achieves the highest neighbor accuracy with minimal beacon overhead.

For the routing performance evaluation, as illustrated in Fig. \ref{f5s3},
the routing update time of all methods exhibits
a positive correlation with the number of nodes.
Given that the time consumed by beacon exchanges is incorporated into the routing table update time, 
the methods with higher beacon exchange frequencies inevitably incur longer routing update durations. 
Owing to its minimal beacon exchange frequency, 
the proposed method achieves the shortest routing table update time.

Furthermore, the results presented in Fig. \ref{f5s2} 
demonstrate that the proposed method possesses excellent scalability in large-scale swarm networks.
As the number of nodes increases from 20 to 80,
the proposed method not only maintains perfect neighbor discovery accuracy but also retains the minimum updating delay of routing.

\subsection{Sensing-and-Communication-Enhanced Control}
We evaluate the performance of collision avoidance algorithm
to verify sensing-and-communication-enhanced control proposed in Section~\ref{control}. 
The simulation parameters are configured as follows.
\begin{itemize}
    \item Range of 3D environment: [300, 300, 100] m.
    \item Dynamic obstacle radius: 20-60 m.
    \item Maximum velocity of UAV: 26 m/s.
    \item Number of iterations: 1000.
\end{itemize}
The UAV is capable of navigating from the designated start point to the destination point 
while ensuring collision avoidance throughout the flight.

\begin{figure}[t]
    \centering
 	\subfloat[]{
		\includegraphics[width=0.5\linewidth]{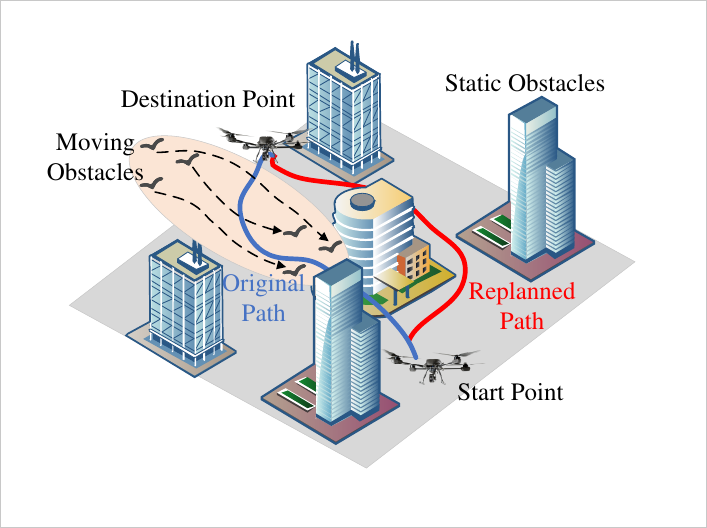} 
		\label{fig7} 
	}
	\subfloat[]{
		\includegraphics[width=0.5\linewidth]{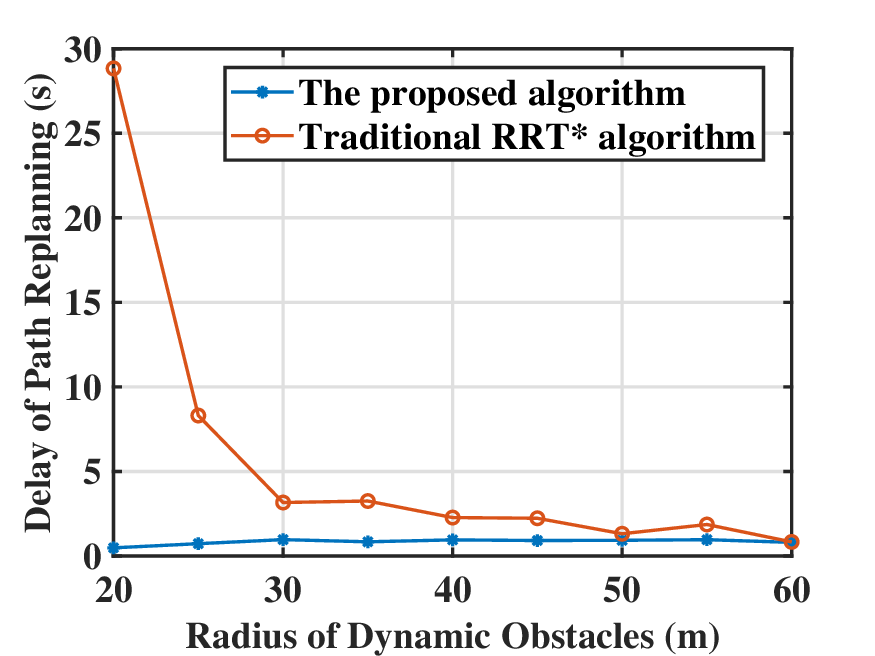}
		\label{fig7b} 
	}
	\caption{Performance evaluation of ISCC-enabled control for UAV swarms: 
                    (a) the path replanning scenario;
                    (b) the relation between path replanning delay and dynamic obstacle radius.}
	\label{fig5}
\end{figure}

As illustrated in Fig. \ref{fig7},
upon predicting a potential collision with a dynamic obstacle,
the UAV first performs an emergency braking maneuver and
then initiates path replanning from its new position after braking,
generating an updated obstacle-avoiding trajectory
(depicted by the red curve). 
Furthermore, Fig. \ref{fig7b} illustrates the correlation
between path replanning delay and the radius of dynamic obstacles,
where the latter is directly determined by sensing accuracy.
By leveraging pre-explored path information to eliminate redundant search operations,
the proposed algorithm achieves lower replanning delay than
the conventional rapidly-exploring random tree* (RRT*) method.
Moreover, as sensing accuracy improves (i.e., as the radius of dynamic obstacles decreases), 
the proposed algorithm exhibits only a marginal reduction in path replanning delay
with the assistance of sensing information, 
whereas the conventional RRT* algorithm experiences a substantial increase in path replanning delay.
These results validate that sensing information
can effectively enhance the control performance of UAVs in complex environments.

\section{Conclusion}
To improve the agility of UAV swarms in disaster relief, aerial BS, 
and logistics transportation scenarios,
we propose an ISCC framework to realize the complementary enhancement of sensing, 
communication and control.
Tailored to scenarios and requirements,
we introduce communication-and-control-enhanced sensing for disaster relief scenario, 
sensing-and-control-enhanced communication for aerial BS scenario, 
and sensing-and-communication-enhanced control for logistics transportation scenario. 
Performance evaluations validate that ISCC framework exhibits superior performance over existing approaches,
which is promising to construct agile UAV swarms empowering low-altitude economy.

\bibliographystyle{IEEEtran}
\bibliography{reference}

\section*{Biography}
\textbf{Zhiqing Wei} (Member, IEEE) received the B.E. and Ph.D. degrees from Beijing University of Posts and Telecommunications (BUPT), Beijing, China, in 2010 and 2015, respectively. He is a Professor with BUPT. His research interest is the performance analysis and optimization of intelligent machine networks. He was granted the Exemplary Reviewer of IEEE WIRELESS COMMUNICATIONS LETTERS in 2017, the Best Paper Award of WCSP 2018 and WCSP 2022. He was the Registration Co-Chair of IEEE/CIC ICCC 2018, the publication Co-Chair of IEEE/CIC ICCC 2019 and IEEE/CIC ICCC 2020.

\textbf{Yucong Du} (Student Member, IEEE) received the B.S. degree and the M.S. degree from Chongqing University of Posts and Telecommunications, Chongqing, China, in 2020 and 2023. He is currently pursuing the Ph.D. degree with the School of Information and Communication Engineering, Beijing University of Posts and Telecommunications (BUPT), Beijing, China. His current research interests include integrated sensing, communication and computing, digital twin network and resource management.

\textbf{Zhiyong Feng} (M'08-SM'15) received her B.E., M.E., and Ph.D. degrees from Beijing University of Posts and Telecommunications (BUPT), Beijing, China. She is a professor at BUPT, and the director of the Key Laboratory of the Universal Wireless Communications, Ministry of Education, China. 
She is a senior member of IEEE, vice chair of the Information and Communication Test Committee of the Chinese Institute of Communications (CIC).
Her main research interests include wireless network architecture design and radio resource management in 5th generation mobile networks (5G), spectrum sensing and dynamic spectrum management in cognitive wireless networks, and universal signal detection and identification.

\textbf{Haotian Liu} (Student Member, IEEE) received the B.E. degree in School of Physic and Electronic Information Engineering, Henan Polytechnic University (HPU) in 2023. 
He is currently pursuing his Ph.D. degree with Beijing University of Posts and Telecommunication (BUPT). 
His research interests include integrated sensing and communication, cooperative sensing, compressed sensing, carrier aggregation.

\textbf{Yanpeng Cui} (Student Member, IEEE) received a B.S. degree from Henan University of Technology, Zhengzhou, China, in 2016, a M.S. degree from Xi'an University of Posts and Telecommunications, China, in 2020, and a Ph.D. degree from Beijing University of Posts and Telecommunications (BUPT), China, in 2024. He is currently a Professor with the School of Information and Engineering, Henan Institute of Science and Technology. 
His research interests include flying ad hoc networks and integrated sensing and communication (ISAC) in unmanned aerial vehicles (UAVs) network.

\textbf{Tao Zhang} received his M.E. degree from Beijing University of Posts and Telecommunications (BUPT). 
He is currently a Chief Wireless Research Fellow and a Professorate level Senior Engineer at the China Unicom Research Institute. His primary research interests lie in mobile communications and wireless networking.

\textbf{Ying Zhou} (Student Member, IEEE) is currently studying for a Ph. D degree in information and communication engineering from Beijing University of Posts and Telecommunications (BUPT). His research interests include wireless communications,  integrated sensing communication and control.

\textbf{Huici Wu} (Member, IEEE) received the Ph.D.
degree from Beijing University of Posts and
Telecommunications (BUPT), Beijing, China, in
2018. From 2016 to 2017, she visited the Broadband
Communications Research (BBCR) Group, University
of Waterloo, Waterloo, ON, Canada. She is now
an Associate Professor at BUPT. Her research interests
are in the area of wireless communications and
networks, with current emphasis on UAV network,
wireless security, authentication, and security game.
\end{document}